\documentstyle[epsf,twocolumn,seceq]{jpsj}
\def\runauthor{Tatsuo. C. {\sc KOBAYASHI} {\it et al}}

\title
{
Pressure-induced Superconductivity in a Ferromagnet UGe$_2$\\
-Resistivity Measurements in Magnetic Field-
}

\author
{ 
Tatsuo C KOBAYASHI, Katsumi HANAZONO$^{1}$, Naoyuki TATEIWA$^{1}$, Kiichi AMAYA$^{1}$, Yoshinori HAGA$^{2}$, Rikio SETTAI$^{3}$ and Yoshichika \=ONUKI$^{3}$
}

\inst
{
Research Center for Materials Science at Extreme Conditions, Osaka University, Toyonaka, Osaka 560-8531, Japan\\
$^1$Department of Physical Science, Grasuate School of Engineering Science, Osaka University, Toyonaka, Osaka 560-8531, Japan\\
$^2$ Advanced Science Research Center, Japan Atomic Energy Research Institute, Tokai-mura, Naka-Gun, Ibaraki, 319-1195, Japan\\
$^3$ Department of Physics, Graduate School of Science, Osaka University, Toyonaka, Osaka 560-0043, Japan\\

}

\recdate
{
\today
}

\abst
{
The electrical resistivity measurements in the magnetic field are carried out on the pressure-induced superconductor UGe$_2$. The superconductivity is observed from 1.06 to 1.44 GPa. The upper critical field of $H_{\rm C2}$ is anisotropic where $H_{\rm C2}(T)$ exhibits positive curvature for $H\, //\, b$ and $c$-axis. The characteristic enhancement of $H_{\rm C2}$ is reconfirmed for $H\, //\, a$-axis. In the temperature and field dependence of resistivity at $P\,>\, P_{\rm C}$ where the ferromagnetic ordering disappears, it is observed that the application of the external field along the {\it a}-axis increases the coefficient of Fermi liquid behavior $AT^{2}$ correspondingly to the metamagnetic transition.
}

\kword
{
, High pressure, UGe$_2$, Ferromagnetism, Superconductivity
}

\begin{document}
\sloppy
\maketitle
\section{Introduction}
 Recently the pressure-induced superconductivity has been found in itinerant ferromagnet UGe$_2$.~\cite{rf:1}  This is unique system where the superconductivity seems to arise from the same electrons that produce band magnetism.

 The resistivity measurement shows that the superconductivity occurs at $T_{\rm SC}$ well below the Curie temperature $T_{\rm C}$ in $1.0\,<\,P\,<1.6$ GPa. It is confirmed by the neutron scattering experiment that the ferromagnetic component of the order is still present at a pressure and temperature where the superconductivity is observed.~\cite{rf:2} The bulk nature of the superconductivity is found by the heat capacity measurement.~\cite{rf:3} These experimental facts verify that the  ferromagnetic ordering and the superconductivity coexist in UGe$_2$.

 It is suggested that the another transition at $T^{*}$ in the ferromagnetic state is related to the appearance of superconductivity.~\cite{rf:2} Namely the $T_{\rm SC}$ shows a maximum at the critical field of $P_{\rm C}^{*}$ where the $T^{*}$ disappears. An unusual reentrant behavior of the superconductivity in the magnetic field along the {\it a}-axis is observed at $P\,=\,1.35$ GPa ($\,> P_{\rm C}^{*}$) where the $T^{*}$ is induced by a magnetic field. These experimental facts are interpreted by considering that the CDW/SDW transition occurs at $T^{*}$. Another characteristic behavior on the transition at $T^{*}$ is an anomalous increment of magnetization. The increment of magnetization reaches 20\% of that above $T^{*}$.~\cite{rf:4, rf:5}

 In this paper, we report the experimental results of the electrical resistivity in the magneic field, focusing the relation between the superconductivity and the disappearance of $T^{*}$ and $T_{\rm C}$.

\section{Experimental}
  A single crystal was grown by the Czochralski pulling method in a tetra-arc furnace as described in ref. 3. The purity of the starting materials was 99.98 \% for U and 99.999 \% for Ge. The ingot was annealed at 800 $^{\circ}$C in high vacuum of 5Å~10$^{-11}$ torr for 7 days. As for the present sample, the residual resistivity $\rho_{0}$ and the residual resistivity ratio RRR (= $\rho_{\rm\, RT}/\rho_{0}$ , $\rho_{\rm \, RT}$: the resistivity at room temperature) were 0.26 $\mu\Omega$cm and 600, respectively, at ambient pressure.

  Pressure was applied by utilizing a indenter cell with a Daphne oil (7373) as a pressure transmitting medium~\cite{rf:6}. The pressure value was determined by the superconducting transition temperatre $T_{\rm SC}$ of lead. The field effect of $T_{\rm SC}$ from the ferromagnetic sample was negligibly small, which was checked at ambient pressure. 
\begin{figure}
 \begin{center}
  \epsfxsize=8.5cm
 \epsfbox{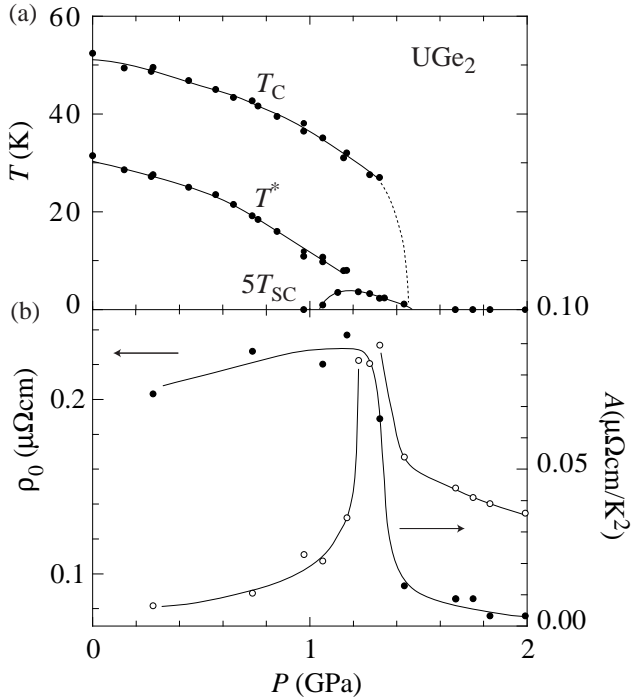}
 \end{center}
\caption{(a) Pressure-temperature phase diagram on UGe$_{2}$. (b) Pressure dependence of $\rho_{0}$ and $A$ in the Fermi liquid behavior of $\rho\,=\,\rho_{0}\,+AT^{2}$.
}
\label{fig:1}
\end{figure} 

\section{Results and discussion}
 The $P$-$T$ phase diagram determined by the electrical resistivity measurement is shown in Fig.1 (a), where $T_{\rm C}$, $T^{*}$ and $T_{\rm SC}$ are determined by the kink, the peak of $d\rho/dT$ and the zero resistance, respectively. Superconductivity is observed from 1.06 to 1.44 GPa. $T_{\rm SC}$ shows the maximum at around $P_{\rm C}^{*}$ = 1.22 GPa where $T^{*}$ disappears. In this experiment, the quantum critical point is considered to be $P_{\rm C} \sim 1.44$ GPa as described below. This critial pressure is slightly different from that reported previously, which may be attributed to the sample dependence or the experimental error of the pressure determination. But it is consistent that superconductivity disappears at around $P_{\rm C}$ and the coefficient $A$ keeps a large value in $P_{\rm C}^{*}\,<\,P\,<\,P_{\rm C}$~\cite{rf:1, rf:2}. The coefficient of Fermi liquid behavior $\rho\,=\,AT^{2}$ increases in the pressure rang of $P_{\rm C}^{*}\,<\,P\,<\,P_{\rm C}$. This indicates the increment of the effective mass of the heavy quasi particles. It is characteristics that there is no increment of $\rho_{0}$ and $A$ at $P_{\rm C} \sim 1.44$ GPa, suggesting that the magnetic-nonmagnetic transition at $P_{\rm C}$ may be a first-order like transition. 

 Figure 2 (a) and (b) show the temperature and field dependence of the resistivity at $P\,=1.67\,$ GPa ($\,>\,P_{\rm C}$). Application of external field along the {\it a}-axis (easy axis) increases rapidly the coefficient $A$, which is considered to be due to the metamagnetic transition from the paramagnetic state at low field to the strongly polarized state at high field.~\cite{rf:7} Further application of the field induces the transition at $T^{*}$ above $H^{*}$ = 7.2 T.~\cite{rf:2} Appearance of $T^{*}$ reduces the coefficient $A$ and increases the residual resistivity. These behavior of $A$ and $\rho_{0}$ corresponds to these pressure dependence at zero field can be seen in Fig.1. At $P$ = 1.44 GPa, the critical field $H_{\rm m}$ exists at zero field, which indicates $P_{\rm C} \sim 1.44$ GPa in the present sample. At $P\,=\,1.22$ GPa, both $T^{*}$ and $H^{*}$ can not be identified in the field and temperature dependences of resistivity, indicating that the critial pressure $P_{\rm C}^{*}$ is close to 1.22 GPa.
\begin{figure}
 \begin{center}
  \epsfxsize=8cm
 \epsfbox{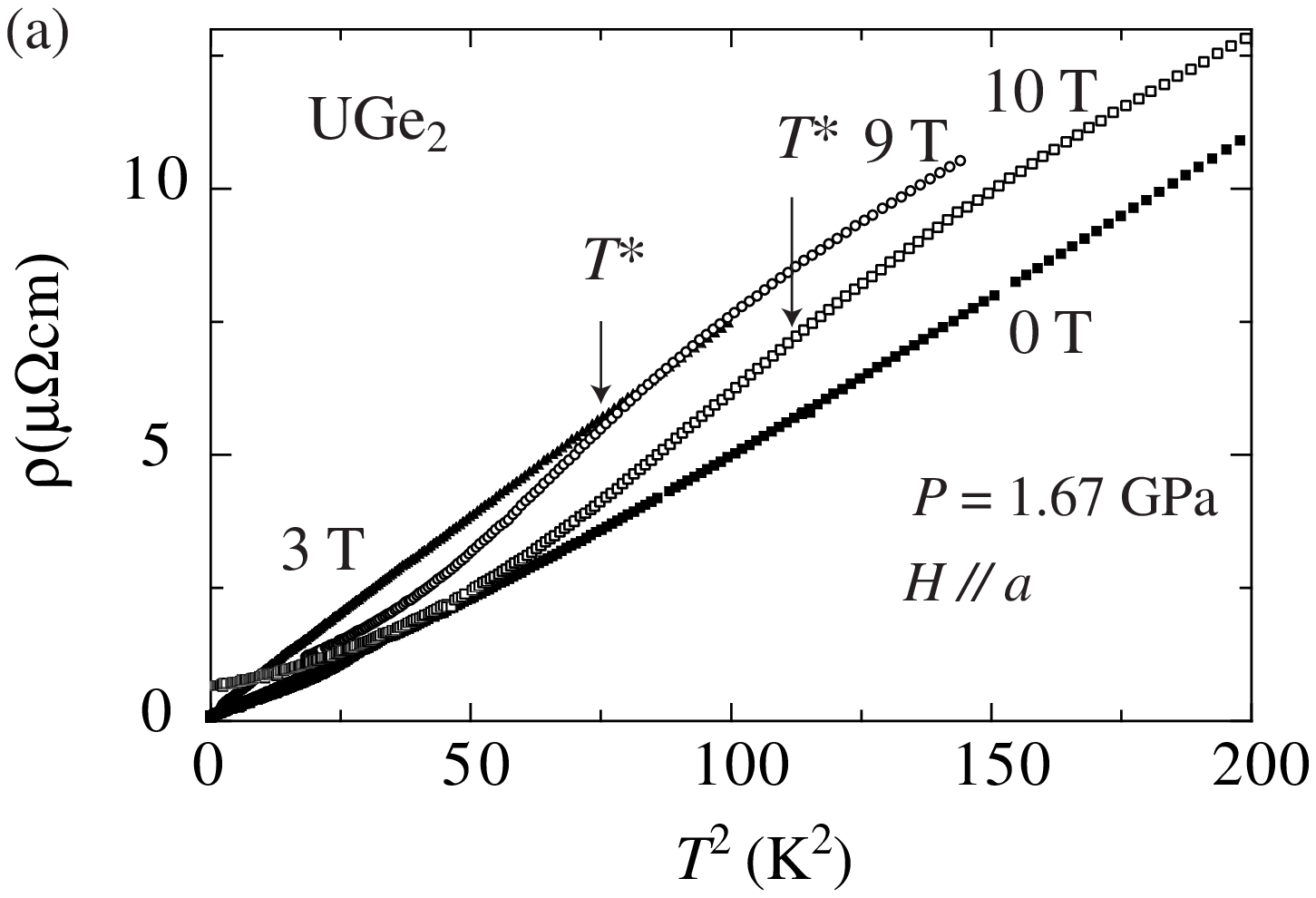}
 \end{center}
\end{figure} 
\begin{figure}
 \begin{center}
  \epsfxsize=8cm
 \epsfbox{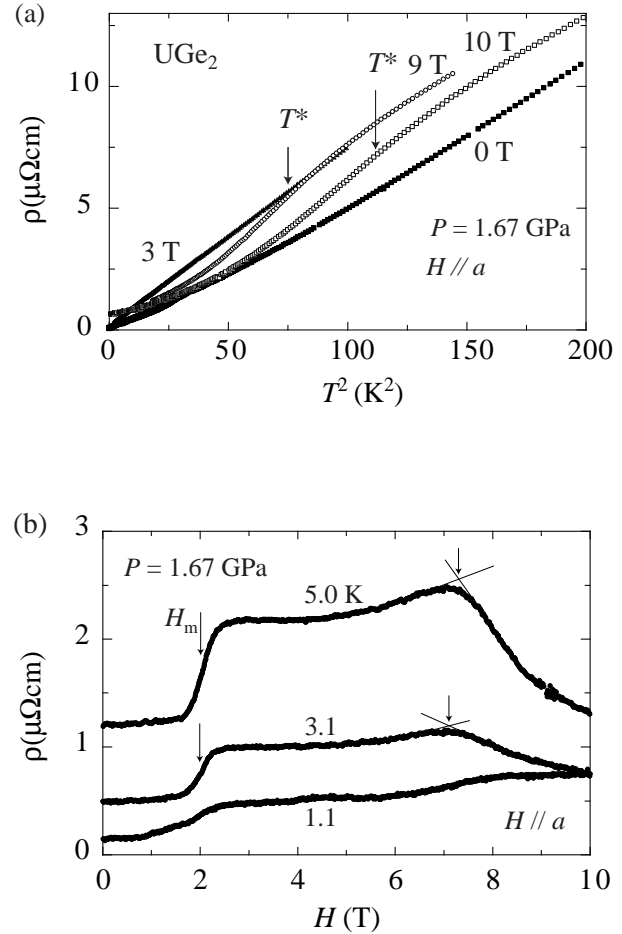}
 \end{center}
\caption{(a) Temperature dependence of the resistivity in magnetic field parallel to the {\it a}-axis at 1.67 GPa ($\, >\,P_{\rm C}$). (b) Field dependence of the resistivity at same pressure.
}
\label{fig:2}
\end{figure} 
\begin{figure}
 \begin{center}
  \epsfxsize=8cm
 \epsfbox{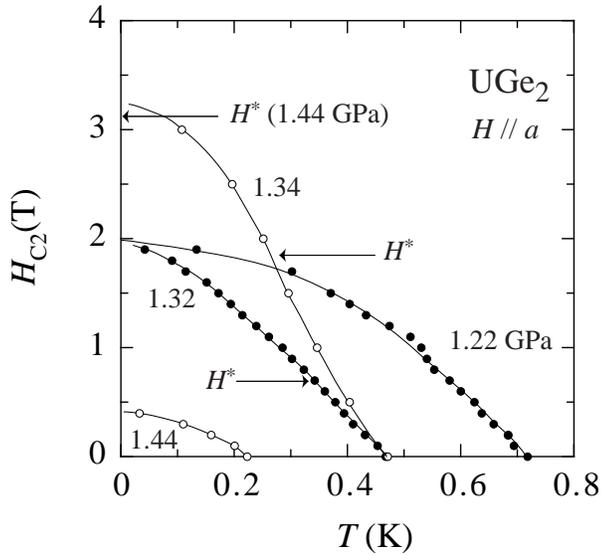}
 \end{center}
\caption{ Superconducting $H-T$ phase diagram at several pressures. External field is applied parallel to the {\it a}-axis (easy axis).
}
\label{fig:3}
\end{figure}
\begin{figure}
 \begin{center}
  \epsfxsize=8.5cm
 \epsfbox{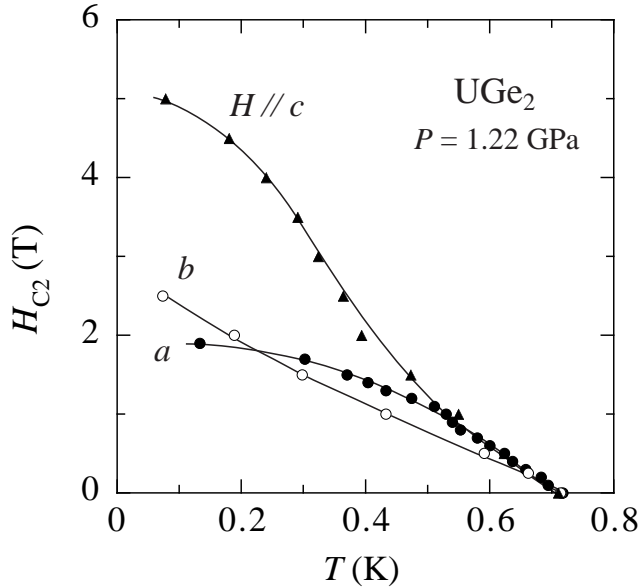}
 \end{center}
\caption{ Anisotropy of the superconducting $H-T$ phase diagram.
}
\label{fig:4}
\end{figure}  

 The superconducting $H-T$ phase diagrams for $H\, //\, a$-axis at several pressures are shown in Fig. 3. The enhancement of the upper critical field $H_{\rm C2}$ is reconfirmed at $P\,=\,1.34$ GPa by tuning $H^{*}$ at low temperature to 2.0 T where the reentrant behavior of superconductivity has been observed in ref. 2. The critical fields $H^{*}$ at each pressure are shown in Fig. 4. The upper critical fields $H_{\rm C2}$ are very sensitive to the critical field $H^{*}$. Watanabe {\it et al.} developed the microscopic theory where the CDW/SDW fluctuation enhances $T_{\rm SC}$ and reproduces qualitatively anomalous superconducting $H-T$ phase diagram.~\cite{rf:8} 

  Figure 4 shows the field dependence of $T_{\rm SC}$ at 1.22 GPa where $T_{\rm SC}$ shows a maximum. The initial slope of $-dH_{\rm C2}/dT$ is about 5.3 T/K for all directions while the upper critical field $H_{\rm C2}$ at the lowest temperature is anisotropic. Here $H_{\rm C2} (T)$ exhibits anomalous positive curvature for $H\, //\, b$ and {\it c}, which is similar with the case of the heavy fermion superconductor UBe$_{13}$~\cite{rf:9}. Similar results of the anisotropic $H_{\rm C2}$ are independently obtained by Sheikin {\it et al.}~\cite{rf:8}

\section{Conclusion}
 In the temperature and field dependence of resistivity at $P\,>\,P_{C}$, the rapid increment of the coeffient $A$ correponding to the metamagnetic transition is found. From these measurement the critical pressures is determined as $P_{\rm C}^{*} \sim 1.22$ GPa and $P_{\rm C}\,\sim 1.44$ GPa. The characteristic enhancement of $H_{\rm C2}$ for $H\, //\, a$-axis is reconfirmed. The upper critical field of $H_{\rm C2}$ is anisotropic where $H_{\rm C2} (T)$ exhibits positive curvature for $H\, //\, b$ and {\it c}.

\section{Acknowledgements}
  We are grateful to K. Miyake and S. Watanabe for helpful discussions. This work is supported by the Grant-in-Aid for COE Research (10CE2004) of the Japanese Ministry of Education, Science, Sports, Culture and Technology, and by CREST of Japan Science and Technology Corporation.

\end{document}